\documentstyle[preprint,aps,epsf]{revtex}                  
\tightenlines
\begin{document}
\pagestyle{empty}                                      
\preprint{
\font\fortssbx=cmssbx10 scaled \magstep2
\hbox to \hsize{
\hfill$\raise .5cm\vtop{
                \hbox{NCTU-HEP-9902}}$}
}
\draft
\vfill
\title{Neutrino-photon scattering and its crossed processes
in a background magnetic field }

\author{Tzuu-Kang Chyi$^a$, Chien-Wen Hwang$^a$, W. F.  Kao$^a$, 
Guey-Lin
Lin$^a$, Kin-Wang Ng$^b$, Jie-Jun Tseng$^a$} %
\address{$^a$\rm Institute of Physics, National Chiao-Tung University,
Hsinchu, Taiwan \rm}
\address{$^b$\rm Institute of Physics, Academia Sinica, Taipei, Taiwan}
\date{\today}
%
%
\vfill
\maketitle
\begin{abstract}
We study the neutrino-photon processes such as
$\gamma\gamma\to \nu\bar{\nu}$,
$\nu\gamma\to \nu\gamma$, and $\nu\bar{\nu}
\to \gamma\gamma$ in a background magnetic field smaller than
the critical magnetic field $B_c\equiv m_e^2/e$.
Using Schwinger's formalism, we extract
leading magnetic-field contributions to the above processes.
Our result is valid throughout the kinematic regime where both
neutrino and
photon energies are significantly smaller than $m_W$.
We briefly discuss the astrophysical implications of our result.
\end{abstract}
%
%
%
%
\pagestyle{plain}
The relevance of neutrino-photon interactions in astrophysics and
cosmology has been
studied extensively\cite{REV}. For example,
the plasmon decay $\gamma^*\to \nu\bar{\nu}$
in horizontal branch stars and red giants
leads to a strong constraint on the
neutrino
magnetic-moment\cite{PLASMON}.
Similarly, the
decay process
$\nu'\to \nu\gamma$ was also calculated\cite{PW}, and its
partial width has been constrained by various astrophysical
observations\cite{REV}.
It is natural to ask whether the two-photon processes such as
the scatterings $\gamma\gamma\to \nu\bar{\nu}$,
$\nu\gamma\to \nu\gamma$ or the decay $\nu'\to \nu \gamma\gamma$
are also relevant in
astrophysics and cosmology. It turns out that,
due to the left-handedness of the weak interaction, the
$O(G_F)$-contributions to the amplitudes of
the above processes are proportional to the mass of the neutrino\cite{MGM}.
Hence the resulting cross
sections or decay rates are very suppressed.
On the other hand, similar processes
involving three
photons such as $\gamma\gamma\to \nu\bar{\nu}\gamma$ or $\nu\gamma\to
\nu\gamma\gamma$ are not suppressed by the same
mechanism\cite{DR}.
Consequently, one expects that
the cross sections for $\gamma\gamma \to \nu\bar{\nu}$ and
its crossed
processes should be enhanced
under a strong background magnetic field. In fact, under a
background magnetic field $B$, the cross section
$\sigma(\gamma\gamma \to \nu\bar{\nu})$ with photon energy 
$E_{\gamma}\ll m_e$
is enhanced by a factor
$(m_W/m_e)^4(B/B_c)^2$\cite{SHA} as compared to its
counterpart in the vacuum, where $m_W$ and
$m_e$ are the $W$ boson and the electron masses respectively; $B_c\equiv
m_e^2/e$ is the critical magnetic field.

The previous calculation on $\gamma\gamma\to \nu\bar{\nu}$\cite{SHA}
applies an effective Lagrangian for
$\gamma\gamma\to \nu\bar{\nu}\gamma$\cite{DR} and replaces
one of the external photon with the classical magnetic field.
It is clear that such an
approach is
valid only
in the limit that $E_{\gamma}, \ E_{\nu}\ll m_e$.
In this work, we shall extend the previous analysis by studying the
processes
$\gamma\gamma\to \nu\bar{\nu}$
, $\gamma\nu\to \gamma\nu$ and $\nu\bar{\nu}\to
\gamma\gamma$ with $E_{\gamma}$ and $E_{\nu}$
larger than $m_e$ but still
considerably smaller than $m_W$. This generalization is motivated
by the fact that the above processes may take place in stars with
temperatures higher than $m_e$. In this case, the effective-Lagrangian
approach is no longer appropriate.

Let us begin with the process
$\gamma\gamma\to \nu\bar{\nu}$ in
a background magnetic field. For convenience,
the cross section of this process is denoted
as $\sigma_B(\gamma\gamma\to \nu\bar{\nu})$. The relevant
Feynman diagram is depicted in
Fig. 1. The effective four-fermion
interactions between leptons and neutrinos can be written as
\begin{equation}
{\cal L}=-{G_F\over 
\sqrt{2}}\left(\bar{\nu}_l\gamma_{\alpha}(1-\gamma_5)
\nu_l\right)\left(\bar{e}\gamma^{\alpha}(g_V-g_A\gamma_5)
e\right),
\end{equation}
where $g_V=1/2+2\sin^2 \theta_w$ and $g_A=1/2$ for $l=e$;
$g_V=-1/2+2\sin^2 \theta_w$ and $g_A=-1/2$
for $l=\mu,\tau$. We should remark that the contribution due to
$g_A$ is proportional to the neutrino mass in the limit of vanishing
magnetic field. 
At $O(B)$ in the limit $B\ll B_c$, 
it gives no contribution to the
amplitude by the charge
conjugation invariance. Therefore we shall neglect
the contribution by $g_A$. Likewise, we shall also neglect contributions
by $g_V$ for $l=\mu,\tau$, since $-1/2+2\sin^2 \theta_w=0.04\ll 1$.
The amplitude for $\gamma(k_1)\gamma(k_2)\to
\bar{\nu}(p_1)\nu(p_2)$ in a background
magnetic field reads:
\begin{eqnarray}
M&=&{G_Fg_V\over \sqrt{2}}4\pi\alpha\bar{u}(p_2)
\gamma^{\rho}(1-\gamma_5)v(p_1)\nonumber \\
&\times&
\int d^4V d^4W tr\left[\gamma_{\rho}{\cal G}(W)\gamma_{\nu}
{\cal G}(-V-W)\gamma_{\mu}{\cal G}(V)\right]\epsilon^{\mu}(k_1)
\epsilon^{\nu}(k_2)\nonumber \\
&\times& \exp({-ie\over 2}V^{\lambda}F_{\lambda\kappa}W^{\kappa})
\exp\left(i(k_1V-k_2W)\right)+(k_1, \mu\longleftrightarrow k_2, \nu)
\label{amp},
\end{eqnarray}
where $V=z-x$, and $W=x-y$; $\epsilon(k_1)$ and $\epsilon(k_2)$
are polarization vectors of the photons; ${\cal G}(W)\equiv
{\cal G}(x-y)$ is a part of the full electron propagator $G(x,y)$
which has the
following form under a constant magnetic field\cite{SCH}
$G(x,y)=\Phi(x,y){\cal G}(x-y)$,
with
$\Phi (x,y)= \exp\left\{ie\int^x_{y} d\xi^\mu \left[A_\mu +
{1\over{2}}F_{\mu\nu} (\xi-y)^{\nu}\right]\right\}$ 
and
\begin{eqnarray}
{\cal G}(x-y)\equiv{\cal G}(W)&=&-i(4\pi)^{-2}
\int_0^{\infty}{ds\over s^2}{eBs\over \sin(eBs)}
\exp(- im_e^2s +ieBs\sigma_3)\nonumber \\
&\times& \exp\left[{-i\over 4s}(W_{\parallel}^2
+eBs\cot(eBs)W_{\bot}^2)\right] \nonumber \\
&\times& \left[m_e+{1\over 2s}\left(\gamma\cdot W_{\parallel}+
{eBs\over \sin(eBs)}
e^{-ieBs\sigma_3}\gamma\cdot W_{\bot}\right)\right],\label{green}
\end{eqnarray}
where $W^{\mu}=W^{\mu}_{\parallel}+
W^{\mu}_{\bot}$ with $W_{\bot}^0=0$ and $\vec{W}_{\bot}
\cdot
\vec{B}=0$, and $\sigma_3=$ \boldmath$\left(
\begin{array}{cc}
 \sigma_3 &  0 \\
 0 & \sigma_3
\end{array}
\right)$ with \boldmath{ $\sigma_3$} the third 
Pauli
matrix. 
\unboldmath
We note that the overall phase $\Phi(x,y)$ breaks the translation 
invariance,
which results from the existence of a constant magnetic field.
The total phase of the three electron-propagators is summarized in the
factor $\exp(-{ie\over 2}V^{\lambda}F_{\lambda\kappa}W^{\kappa})$.
This phase is easily obtained by realizing that
the combination $d\xi^{\mu}(A_{\mu}+{1\over 2}F_{\mu\nu}\xi^{\nu})\equiv
{\cal A}$ in the expression for $\Phi(x,y)$ is an exact
form which satisfies ${\cal A}=d\omega$. Therefore the
integration of ${\cal A}$ around a closed loop vanishes. The total phase 
of
the electron propagators is then given by
\begin{eqnarray}
\Phi(x,y)\cdot \Phi(y,z) \cdot \Phi(z,x)&=&
\exp(-{ie\over 2}V^{\mu}F_{\mu\nu}W^{\nu}).
\end{eqnarray}
For a constant magnetic field along $+z$ direction, 
we have $F_{12}=-F_{21}=B$ while other components of $F_{\mu\nu}$
vanish. 

At this stage, the calculation of $M$ remains nontrivial since the
function ${\cal G}$ as given by Eq. (\ref{green})
is complicated. To find a simplification for ${\cal G}$, we
go to the momentum space, which amounts to writing
${\cal G}(x,y)=\int {d^4p\over {(2 \pi)^4}} e^{-ip(x-y)} {\cal G}(p),$
with
\begin{eqnarray}
{\cal G}(p)
&=&\int^\infty_0 {ds\over {\cos(eBs)}}\exp\left[-is\left(m_e^2-p^2_
{\parallel}-
{\tan(eBs)\over{eBs}} p^2_{\bot}\right)\right]\nonumber \\
&&~~~~~~~\times \left[e^{ieBs\sigma_3}(m_e+\gamma\cdot 
p_{\parallel})+{\gamma
\cdot p_{\bot}\over {\cos(eBs)}}\right].\label{Gp}
\end{eqnarray}
It is useful to write ${\cal G}$ in terms of Landau levels\cite{CEO}
\begin{equation}
{\cal G}(p)=\sum_{n=0}^{\infty}{-id_n(\alpha)(\gamma\cdot
p_{\parallel}+m_e+\gamma\cdot p_{\bot} )+d_n^{\prime}(\alpha)
\gamma_1\gamma_2(m_e+\gamma\cdot p_{\parallel})\over p_L^2+2neB}
+i{\gamma\cdot p_{\bot}\over p_{\bot}^2},
\label{landau}
\end{equation}
where $p_L^2=m_e^2-p_{\parallel}^2$, $\alpha=-p_{\bot}^2/eB$, and
$d_n(\alpha)=(-1)^ne^{-\alpha}(L_n(2\alpha)-L_{n-1}(2\alpha))$
with $L_n$ the Laguerre polynomials.
As indicated by Eq. (\ref{landau}), the $B$ dependence of ${\cal G}(p)$
resides in $d_n(\alpha)$, $d'_n(\alpha)$, 
and the propagator $1/(p_L^2+2neB)$. 
For $B\ll B_c$, 
the propagator
${\cal G}$ and the phase factor 
$\exp({-ie\over 2}V^{\lambda}F_{\lambda\kappa}W^{\kappa})$ can be expanded 
in powers of $eB$.  
To the linear order in $eB$, we have\cite{CTAS}
\begin{equation}
{\cal G}(p)=i{\gamma\cdot p +m_e\over p^2-m_e^2+i\epsilon}-
{\gamma^1\gamma^2(m_e+\gamma\cdot p_{\parallel})
\over (p^2-m_e^2+i\epsilon)^2}eB +O(e^2B^2),
\label{series}
\end{equation}
and $\exp({-ie\over 2}V^{\lambda}F_{\lambda\kappa}W^{\kappa})$
$= 1-{ie\over 2}\left(V^{\lambda}F_{\lambda\kappa}W^{\kappa}\right)
+O(e^2B^2)$.
The above expansions can be used to compute the amplitude 
$M$ in powers of $eB$. Indeed, by dimensional analysis, any 
given power of $eB$ in the expansion of $M$ is accompanied by
an equal power of $1/m_e^2$(for $m_e >p$) or $1/p^2$(for 
$p> m_e$) with $p$ the typical 
energy scale of external particles. Clearly, for $B\ll B_c\equiv
m_e^2/e$, both $(eB/m_e^2)^n$ and $(eB/p^2)^n$(applicable 
when $p> m_e$) are 
much smaller than unity.  

From Eqs. (\ref{amp}), (\ref{series}) and the expansion of 
the phase factor,
the amplitude $M$ to
the linear order in $eB$ is
\begin{equation}
M={G_Fg_V\over \sqrt{2}}{e\alpha\over 4\pi}
\bar{u}(p_2)\gamma_{\rho}(1-\gamma_5)v(p_1)J^{\rho},\label{amp2}
\end{equation}
with
\begin{eqnarray}
J^{\rho}&=&C_1\left[(\epsilon_1 
F\epsilon_2)(k_1^{\rho}-k_2^{\rho})\right]
+C_2\left[(\epsilon_1Fk_1)(k_1\cdot \epsilon_2)k_2^{\rho}
+(\epsilon_1Fk_1)(k_1\cdot \epsilon_2)k_2^{\rho}\right]\nonumber \\
&+&C_3\left[(\epsilon_1Fk_1)\epsilon_2^{\rho}+
(\epsilon_2Fk_2)\epsilon_1^{\rho}\right]
+C_4\left[(\epsilon_1Fk_2)(k_1\cdot \epsilon_2)k_1^{\rho}
+(\epsilon_2Fk_1)(k_2\cdot \epsilon_1)k_2^{\rho}\right]\nonumber \\
&+& C_5\left[(\epsilon_1Fk_2)(k_1\cdot \epsilon_2)k_2^{\rho}
+(\epsilon_2Fk_1)(k_2\cdot \epsilon_1)k_1^{\rho}\right]
+C_6\left[(\epsilon_1Fk_2)\epsilon_2^{\rho}+
(\epsilon_2Fk_1)\epsilon_1^{\rho}\right]\nonumber \\
&+&C_7(k_2\cdot \epsilon_1)(k_1\cdot \epsilon_2)
\left[(Fk_1)^{\rho}+(Fk_2)^{\rho}\right]
+C_8(\epsilon_1\cdot \epsilon_2)
\left[(Fk_1)^{\rho}+(Fk_2)^{\rho}\right]\nonumber \\
&+&C_9\left[(k_1Fk_2)(\epsilon_1\cdot \epsilon_2)
(k_1^{\rho}-k_2^{\rho})\right]
+C_{10}\left[(k_1Fk_2)(k_2\epsilon_1)(k_1\epsilon_2)
(k_1^{\rho}-k_2^{\rho})\right]\nonumber \\
&+&C_{11}\left[(k_1Fk_2)(k_2\cdot \epsilon_1\epsilon_2^{\rho}
+k_1\cdot \epsilon_2\epsilon_1^{\rho})\right]\label{current},
\end{eqnarray}
where $C_1, C_2\cdots , C_{11}$ are 
linear combinations of the 
integrals
$I[a,b,c]=\int_0^1 dx\int_0^x dy{x^by^{a-b}\over (1-txy-i\varepsilon)^c}$
with $t=2k_1\cdot k_2/ m_e^2$.
The detailed structures of these coefficients will be
presented elsewhere\cite{CTAS}.
We have checked our result by taking the limit $2k_1\cdot k_2/m_e^2 \ll
1$. It agrees with the result of Ref.\cite{SHA}, which is obtained
via the effective-Lagrangian approach. 

From Eqs. (\ref{amp2}) and
(\ref{current}), we can calculate the cross section for $\gamma\gamma
\to \nu\bar{\nu}$ in a background magnetic field.
For simplicity, let us
take the momenta of incoming photons to be
along $+z$ and $-z$ directions respectively, with equal magnitudes.
The result
for $\sigma_B(\gamma\gamma
\to \nu\bar{\nu})$ with $B=0.1 \ B_c$ and $\vec{B}$
perpendicular to the collision axis
is plotted
in Fig. 2. For other relative alignments between
$\vec{B}$ and the collision axis, the cross
section $\sigma_B$ varies by no more than
an order of magnitude.
To explore the validity of the effective-Lagrangian approach,
we also plot the cross section
$\sigma^*_B(\gamma\gamma\to \nu\bar{\nu})$ obtained in this method\cite{SHA}.
It is found that $\sigma_B$ and $\sigma^*_B$
agree reasonably well at a small incoming
photon energy ($\omega$), i.e., $\omega/m_e < 0.5$. For
$\omega$ slightly greater than $m_e$, the internal
electron could become on shell, and $\sigma_B$ would dominate over
$\sigma^*_B$ due to
the rescattering effect by $e^+e^-\to
\nu\bar{\nu}$. Such a dominance lasts till
$\omega/m_e=2.2$ where $\sigma^*_B$ begins to overtake $\sigma_B$.
Finally, for comparisons, we also display the $2\to 3$
scattering cross section $\sigma(\gamma\gamma\to \nu\bar{\nu}\gamma)$
obtained in Refs.\cite{DKR,AMP}. For $\omega/m_e < 5$,
this cross section is seen
to be suppressed compared to $\sigma_B(\gamma\gamma\to \nu\bar{\nu})$.
At higher energies, it becomes equally important as the latter.

The stellar energy-loss rate $Q$ due to
$\gamma\gamma\to \nu\bar{\nu}$ in a
background magnetic field has been calculated\cite{SHA}.
We repeat the
calculation using our updated result of $\sigma_B(\gamma\gamma\to
\nu\bar{\nu})$.
The temperature dependencies of $Q$
are listed in Table I. For comparisons, we also list corresponding 
results
obtained from the effective-Lagrangian approach
\cite{COMM}. For temperatures below $0.01$ MeV, the effective-Lagrangian 
approach works very well. On the other
hand, this approach becomes rather inaccurate for temperatures greater
than $1$ MeV. At $T=0.1$ MeV, our exact calculation gives an 
energy-loss
rate almost two orders of magnitude greater than the result from the
effective Lagrangian. Such a behavior can be understood from the
energy dependence of the scattering cross section, as shown in Fig. 2.
It is clear that, for $T=0.1$ MeV, $Q$ must have received
significant contributions from scatterings with
$\omega\approx m_e$. At this energy, the full calculation gives a much 
larger
scattering cross section than the effective Lagrangian does.

By comparing the predictions of the full calculation and
the effective-Lagrangian approach\cite{SHA}, we conclude that
the applicability of the latter to the energy-loss rate
is quite restricted. While the effective Lagrangian works reasonably 
well
with $\omega< \,  0.1 m_e $ , it would give a poor approximation on $Q$
unless $T < \, 0.01 m_e$.

Besides $\gamma\gamma\to \nu\bar{\nu}$, the crossed processes
$\nu(\bar{\nu})\gamma\to \nu(\bar{\nu})\gamma$ and
$\nu\bar{\nu}\to \gamma\gamma$ in a background magnetic field
also play some roles in astrophysics. For example, one expects that 
these
two processes might be relevant for the mean free paths of
supernova neutrinos. In fact, it was recently suggested that\cite{TEP},
for supernova neutrinos, the $2\to 3$ scatterings
$\nu\gamma\to \nu\gamma\gamma$ and $\nu\bar{\nu}\to \gamma\gamma\gamma$
give neutrino mean free paths less than the supernova core radius.
Thus they could affect the supernova dynamics.
Now since the magnetic field inside the supernova core is
typically around $10^{12} G\approx 0.1 B_c$, the cross
sections $\sigma_B(\nu(\bar{\nu})\gamma\to \nu(\bar{\nu})\gamma)$
and $\sigma_B(\nu\bar{\nu}\to \gamma\gamma)$ are expected to be
comparable
to those of $2\to 3$ scattering just mentioned. Hence one might
conclude
that $\nu(\bar{\nu})\gamma\to \nu(\bar{\nu})\gamma$ and
$\nu\bar{\nu}\to \gamma\gamma$ are also relevant for the
supernova dynamics. To examine this statement, one should note that
the result of Ref. \cite{TEP} is based upon
extrapolating the energy dependencies of 
$\sigma(\nu\gamma\to \nu\gamma\gamma)$ and
$\sigma(\nu\bar{\nu}\to \gamma\gamma\gamma)$ obtained 
in the low energy limit $E_{\nu},\  E_{\gamma}\ll m_e$\cite{DR}
to energies greater than few times of $m_e$. 
Naturally, such an extrapolation
is likely to overestimate the cross sections at higher
energies, resulting into
an underestimation of the corresponding neutrino mean free paths.
With this precaution in mind, we shall first compute
$\sigma_B(\nu(\bar{\nu})\gamma\to \nu(\bar{\nu})\gamma)$,
$\sigma_B(\nu\bar{\nu}\to \gamma\gamma)$ and their
corresponding neutrino mean free paths. 
The mean free paths due to $2\to 3$ scatterings can then be
easily inferred. Hence the results of Ref. \cite{TEP} can be checked.

The amplitude for $\gamma\nu\to \gamma\nu$ can be inferred from
Eqs. (\ref{amp2}) and (\ref{current}) with $v(p_1)\to u(p_1)$ and
$k_1\to -k_1$. It is worth noting that, unlike $\gamma
\gamma\to \nu\bar{\nu}$, this process develops no imaginary part
since there are no intermediate states available for the rescattering.
In Fig. 3, we show
the cross sections of $\gamma\nu\to \gamma\nu$ as a function of
the incoming photon energy in the center of momentum frame.
We have presented two cross sections with the magnetic field parallel and
perpendicular to the collision axis respectively.
For most incoming photon energies, these two cross sections, denoted as
$\sigma_B(\gamma\nu\to \gamma\nu)_{\parallel}$ and
$\sigma_B(\gamma\nu\to \gamma\nu)_{\bot}$
respectively,
differ by no more than an order of magnitude.
For comparisons,
we also display the $2\to 3$ scattering cross section
$\sigma(\gamma\nu\to \gamma\gamma\nu)$\cite{DKR,AMP}.
It is clear that the $2\to 2$ cross section with $B=0.1\, B_c$ is
significantly greater than the $2\to 3$ cross section for
$\omega \leq m_e$. These two cross sections become
comparable for $\omega > m_e$. At $\omega=m_e$, we have, for example,
$\sigma_B(
\gamma\nu\to \gamma\nu)_{\parallel}=1.6\times 10^{-53}\, {\rm cm}^2$, and
$\sigma(\gamma\nu\to \gamma\gamma\nu)=1\times 10^{-55}\, {\rm cm}^2$.
For $\omega=50 \ m_e$, both cross sections reach to roughly
$10^{-49}\, {\rm cm}^2$. We also note that $\sigma_B(\gamma\nu
\to \gamma\nu)_{\parallel(\bot)}$
is a smooth function of $\omega$
for the energy range considered here. In fact, the cross section
maintains such a smooth behavior
until $\omega$ approaches to $m_W$. 

The neutrino mean free path implied
by the above $\nu\gamma$ scattering, which we denote as
$\lambda_1$, can be calculated using
$\lambda_1={1 /n_{\gamma}\sigma_{\nu\gamma}}$,
where $n_{\gamma}$ is the photon number density, and
$\sigma_{\nu\gamma}$ is the average neutrino-photon scattering cross section.
Since we simply concern the order of magnitude of
$\lambda_1$, we shall assume the momenta of the photon and the
neutrino
to be along $+z$ and $-z$ directions respectively, while $\vec{B}$
is taken to be parallel to the collision axis. 
The neutrino mean free paths for different neutrino energies
are summarized in Table II. 
For $T=20$ MeV, 
$E_{\nu}=20$
MeV, $\mu_{\nu}=0$(a vanishing neutrino chemical potential) and 
$B=0.1\,
B_c$,
we find $\lambda_1=3\times 10^{14}$ cm, which is much
greater than $10^6$ cm,
the supernova core radius. The neutrino mean free path decreases to 
$4\times
10^{13}$ cm for $E_{\nu}=50$ MeV, and increases to $1\times 10^{17}$ 
cm
for $E_{\nu}=1$ MeV. 
Clearly the photon medium in the supernova is
transparent to the neutrino as far as the scattering $\nu\gamma\to
\nu\gamma$
is concerned. Hence this process is not expected to affect the
supernova dynamics.
Furthermore, since the cross section $\sigma(\nu\gamma\to
\nu\gamma\gamma)$ is at most comparable to $\sigma_B(\nu\gamma\to
\nu\gamma)_{\parallel(\bot)}$, the neutrino mean free path implied by
the former process should also be much greater than the supernova core
radius. This is in a sharp contrast to the small neutrino mean free path
($\lambda_{\nu\gamma\to
\nu\gamma\gamma}\approx 10^{-3}$ cm for $T=20$ MeV, $E_{\nu}=20$
MeV, and $\mu_{\nu}=0$) obtained in
Ref. \cite{TEP}. This discrepancy confirms that the
extrapolation performed in Ref.\cite{TEP} indeed 
underestimates the mean free paths of high energy neutrinos
with $E_{\nu} > m_e$. 

Now let us turn to the last process,
$\nu\bar{\nu}\to \gamma\gamma$ in a background magnetic field.
This process behaves rather similarly as
the reversed process $\gamma\gamma\to \nu\bar{\nu}$ discussed before.
The scattering cross section $\sigma_B(\nu\bar{\nu}\to \gamma\gamma)_
{\parallel(\bot)}$
is depicted in Fig. 4.
For comparisons, the corresponding
$2\to 3$ cross section $\sigma(\nu\bar{\nu}\to \gamma\gamma\gamma)$
\cite{DKR,AMP} is also shown.
One can see that $\sigma_B(\nu\bar{\nu}\to 
\gamma\gamma)_{\parallel(\bot)}$
peaks locally
in the vicinity of
$\omega=m_e$ where the threshold effect of electron pair-production
emerges.
Furthermore, from
$\omega=0.1 \, m_e$ to
$\omega=m_e$, the $2\to 2$ cross section dominates the $2\to 3$
cross section by a few orders of magnitude.
The two curves cross at $\omega\approx 5 \ m_e$, at which point
the $2\to 3$ process begins to dominate. At $\omega=m_e$,
$B=0.1 \ B_c$,
we have, for example,
$\sigma_B(\nu\bar{\nu}\to \gamma\gamma)_{\parallel}=
10^{-49}{\rm cm}^2$,
and $\sigma(\nu\bar{\nu}\to \gamma\gamma\gamma)=
1.5\times 10^{-53}{\rm cm}^2$\cite{DKR,AMP}. The former cross section 
becomes
$3\times 10^{-50}{\rm cm}^2$ at $\omega=50 \ m_e$ while the latter
cross section is roughly an order of magnitude larger.
The neutrino mean free path due to $\nu\bar{\nu}\to \gamma\gamma$,
which we denote as $\lambda_2$,
can be calculated using 
$\lambda_2={1 /n_{\bar{\nu}}\sigma_{\nu\bar{\nu}}}$,
where $n_{\bar{\nu}}$ is the number density of the antineutrino, and
$\sigma_{\nu\bar{\nu}}$ is the average  
cross section of $\nu\bar{\nu}\to \gamma\gamma$. The results on $\lambda_2$
for different neutrino energies are listed in Table II.
For $T=20$ MeV,  $E_{\nu}=20$
MeV, $\mu_{\nu}=0$ and $B=0.1\, B_c$, we find
$\lambda_2=5\times 10^{16}$ cm.
The neutrino mean free path decreases to $3\times
10^{15}$ cm for $E_{\nu}=50$ MeV, and increases to $3\times 10^{18}$ 
cm
for $E_{\nu}=1$ MeV. Once again, the above neutrino mean free paths
are all much greater than the supernova core radius. Furthermore, by
comparing $\sigma(\nu\bar{\nu}\to \gamma\gamma\gamma)$ with
$\sigma_B(\nu\bar{\nu}\to \gamma\gamma)_{\parallel(\bot)}$, we conclude
that the neutrino mean free path 
relevant to the former process should also be much 
greater
than the supernova core radius. This is again in a sharp contrast to a
small neutrino mean free path
($\lambda_{\nu\bar{\nu}\to \gamma\gamma\gamma}\approx 10^{-3}$ cm for 
$T=20$ MeV, $E_{\nu}=20$
MeV, and $\mu_{\nu}=0$) obtained in
Ref. \cite{TEP}. Once more, this discrepancy is due to 
the cross section extrapolation performed in Ref. \cite{TEP}. 

In conclusion, we have illustrated the weak-field expansion
technique for processes occurring in a background magnetic field.
Specifically, we apply this technique to
calculate the cross sections of
$\gamma\gamma \to \nu\bar{\nu}$, $\gamma\nu\to \gamma\nu$, and
$\nu\bar{\nu}\to \gamma\gamma$ under a background magnetic field.
We found that the effective-Lagrangian approach is inappropriate
for computing the stellar energy-loss rate due to
$\gamma\gamma \to \nu\bar{\nu}$, unless the star temperature is
less than $0.01\, m_e$. We also found that the neutrino mean free paths
relevant to $\gamma\nu\to \gamma\nu$ and
$\nu\bar{\nu}\to \gamma\gamma$
in a background magnetic field
are much greater than the supernova
core radius. The same conclusions are reached for the neutrino mean
free paths relevant to
$\gamma\nu\to \gamma\gamma\nu$ and
$\nu\bar{\nu}\to \gamma\gamma\gamma$.
Therefore both neutrino-photon scatterings and neutrino-antineutrino
annihilations
into photons are not expected to affect the supernova dynamics.

\acknowledgments
This work was supported in part by the National Science Council of 
R.O.C.
under the Grant Nos. NSC-88-2112-M-009-001, NSC-88-2112-M-009-002,
and NSC-88-2112-M001-042.

\begin{table}
\caption{The temperature dependence of energy-loss rate(erg/s =
$\rm{cm}^3$) by
$\gamma\gamma\to \nu\bar{\nu}$ in a background magnetic field.
The results given by the effective Lagrangian and our exact calculations
are both listed. We take $B=B_c/10$.}
 \begin{tabular}{|l|c|c|c|c|c|}
  Q/T(MeV)     & 0.001 &0.01 &0.1 &1 &10            \\ \hline
Exact          &$1.2\times 10^{-18}$ &$1.2\times 10^{-5}$
               &$5.6\times 10^9$  & $1.7\times 10^{18}$
               &$3.1\times 10^{26}$                 \\ \hline
Effective      &$1.2\times 10^{-18}$ &$1.2\times 10^{-5}$
               &$1.2\times 10^8$  & $1.2\times 10^{21}$
               &$1.2\times 10^{34}$
\end{tabular}
\end{table}
\begin{table}
\caption{The neutrino mean free paths $\lambda_{1,2}$ relevant to
the scatterings $\nu\gamma\to \nu\gamma$ and 
$\nu\bar{\nu}\to \gamma\gamma$ respectively in a background magnetic field
$B=B_c/10$. We fix $T=20$ MeV, $\mu_{\nu}=0$ and vary the 
neutrino energy $E_{\nu}$ from $0.01$ MeV to $50$ MeV. }
\begin{tabular}{|l|c|c|c|c|c|c|}
$E_{\nu}$(MeV) &0.01 &0.1 &1 &5 &20 &50        \\ \hline
$\lambda_1$ (cm)   &$1\times 10^{22}$ &$2\times 10^{19}$
&$1\times 10^{17}$ &$5\times 10^{15}$ &$3\times 10^{14}$
&$4\times 10^{13}$                              \\ \hline
$\lambda_2$ (cm)   &$2\times 10^{21}$ &$1\times 10^{17}$
&$3\times 10^{18}$ &$1\times 10^{18}$ &$5\times 10^{16}$
&$3\times 10^{15}$                            

\end{tabular}
\end{table}
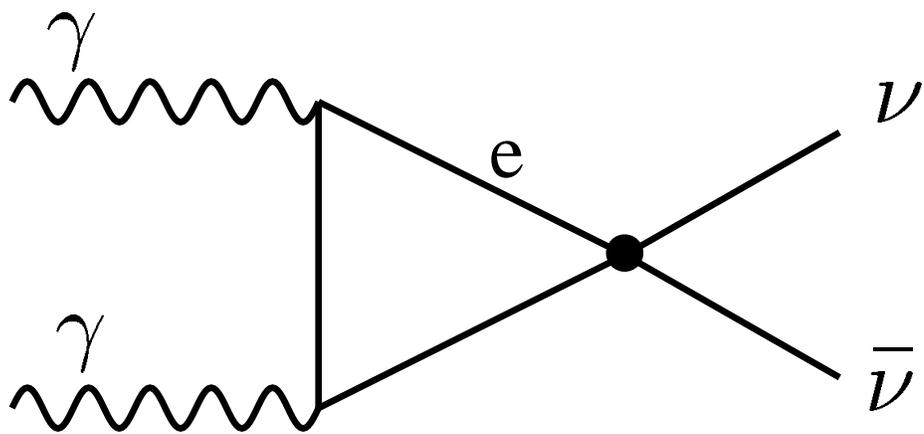
\begin{figure}
\caption{Feynman diagrams contributing to
$\gamma\gamma\to \nu\bar{\nu}$.}
\label{fig1}
\end{figure}
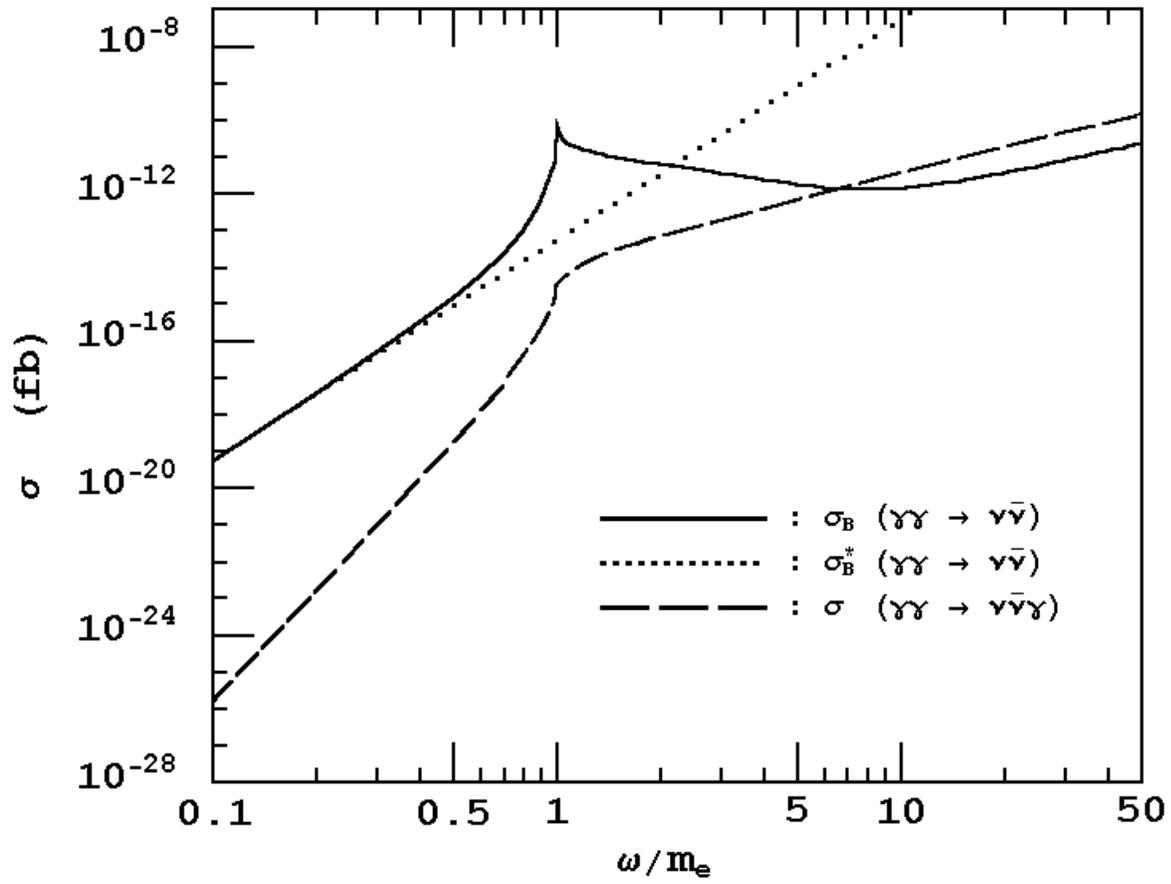
\begin{figure}
\caption{$\sigma_B(\gamma\gamma\to \nu\bar{\nu})$
is the cross section obtained from the exact calculation, while
$\sigma_B^*(\gamma\gamma\to \nu\bar{\nu})$ is obtained from the
effective Lagrangian approach.
The magnetic field direction is taken to
be parallel to the collision
axis.
For comparison, the $2\to 3$ cross section
$\sigma(\gamma\gamma\to \nu\bar{\nu}\gamma)$ is also displayed.}
\label{fig2}
\end{figure}
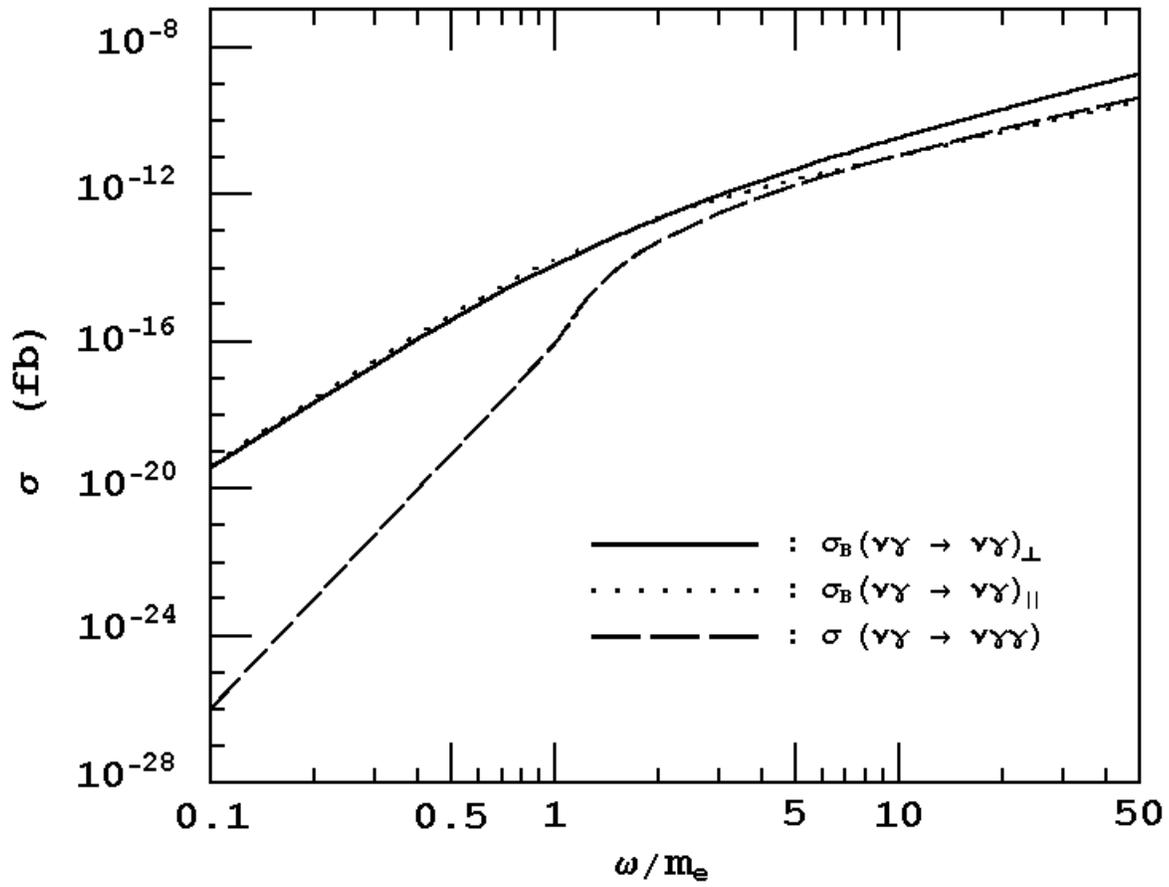
\begin{figure}
\caption{The solid line and the dotted line depict
cross sections $\sigma_B(\gamma\nu\to \gamma\nu)_{\bot}$ and
$\sigma_B(\gamma\nu\to \gamma\nu)_{\parallel}$ respectively.
The dashed line depicts the
cross section $\sigma(\gamma\nu\to \gamma\gamma\nu)$.}
\label{fig3}
\end{figure}
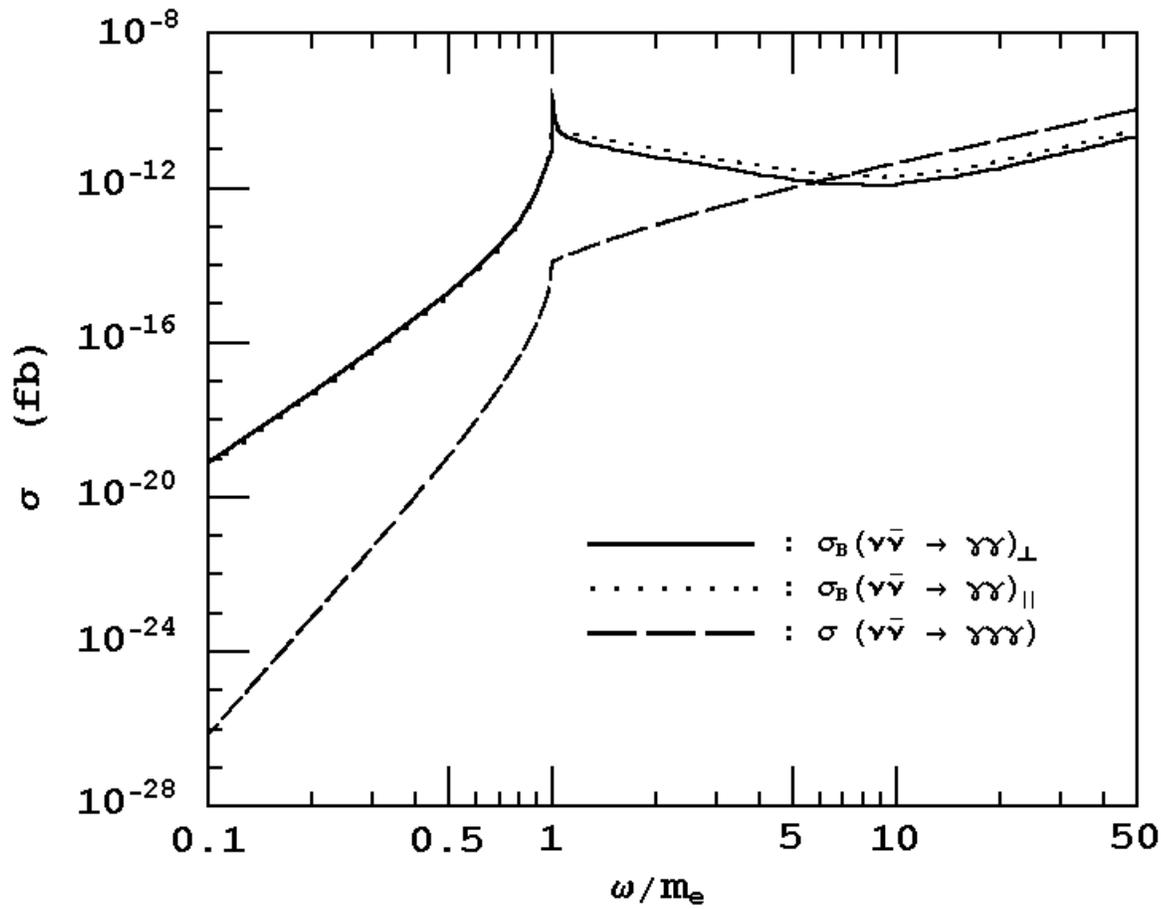
\begin{figure}
\caption{The solid line and the dotted line depict
cross sections $\sigma_B(\nu\bar{\nu}\to \gamma\gamma)_{\bot}$ and
$\sigma_B(\nu\bar{\nu}\to \gamma\gamma)_{\parallel}$ respectively.
The dashed line depicts the
cross section $\sigma(\nu\bar{\nu}\to \gamma\gamma\gamma)$.}
\label{fig4}
\end{figure}


\begin{thebibliography}{99}
\bibitem{COMM}
In the low temperature limit, i.e., $T\ll m_e$, our $Q$ differs
from that of Ref.\cite{SHA}, but the discrepancy is within
one order of magnitude.
\end{thebibliography}
\end{document}